\documentclass[journal=jctcce, manuscript=article, 
                        layout=twocolumn 
                        ]{achemso}

\usepackage{chemformula} 
\usepackage[T1]{fontenc} 
\usepackage{amsmath, amsfonts, amssymb, amsthm} 
\usepackage{bm} 
\usepackage{hyperref} 
\usepackage{nameref}  
\usepackage[version=4]{mhchem} 

\usepackage{soul}  
\soulregister\ref7
\soulregister\cite7
\soulregister\ce7
\soulregister\textrm7

\author{Leonard P. Heinz}
\affiliation[MPINAT]{Department of Theoretical and Computational Biophysics, Max-Planck Institute for Multidisciplinary Sciences, G{\"o}ttingen, Germany}
\alsoaffiliation[priv]{Present address: d-fine GmbH, Frankfurt, Germany}

\author{Helmut Grubm{\"u}ller}
\affiliation[MPINAT]{Department of Theoretical and Computational Biophysics, Max-Planck Institute for Multidisciplinary Sciences, G{\"o}ttingen, Germany}
\email{hgrubmu@gwdg.de}

\newcommand{\kJmol}{kJ$\cdot$mol$^{-1}$}

\title{Why solvent response contributions to solvation free energies are compatible with Ben-Naim's theorem
}

\abbreviations{}

\begin{document}
\setcounter{secnumdepth}{3}

\begin{abstract}
We resolve a seeming paradox arising from a common misinterpretation of Ben-Naim’s theorem, which rests on a pairwise decomposition of the Hamiltonian of a molecular solute/solvent system into pairwise solute-solvent and solvent-solvent interactions. According to this theorem, also the solvation entropy can be decomposed into a solute-solute term and a remaining term that is --- perhaps misleadingly --- referred to as "solvent reorganization entropy". Crucially, the latter equals the average solvent-solvent interaction energy, such that these two solvent-solvent terms do not change the total solvation free energy. This analytical result has often been used to argue that the reorganization of the solvent cannot play a role in the solvation process, and thus to rule out "iceberg"-type solvent shell ordering as a solvation driving force. However, recent calculations based on atomistic simulations of a solvated globular protein and spatially resolved mutual information expansions revealed substantial contributions of many-body solvent correlations to the solvation free energy. Here we resolved this seeming contradiction and illustrate by two examples --- a simple Ising model and a solvated Lennard-Jones particle --- that the solvent reorganization entropy and the actual entropy contribution arising from many-body solvent correlations differ both conceptually and numerically. Whereas the solvent reorganization entropy in fact arises from both solvent-solvent as well as solute-solvent interactions and thus fails to describe what the name suggests, the mutual information expansion permits a straightforward interpretation in terms of the entropy contribution of solvent-solvent correlations to the solvation free energy. 
\end{abstract}

\section{Introduction}
\label{sec:introduction}

The hydrophobic effect is an essential driving force for many processes in nature, such as phase separation, membrane formation\cite{Israelachvili_1976, Devrjes_2004_bilayer_assembly, Maibaum_2004}, or the function and folding of proteins\cite{Chandler_2005, Dias_2010_cold_denaturation}. Despite its significance, the hydrophobic effect is not yet fully understood from first principles, and hence its molecular explanation remains controversial\cite{Hummer_2000}. The early "iceberg hypothesis" by Frank and Evans\cite{iceberg_model}, for example, turned out to be equally popular and controversial. Frank and Evans explained the unfavorable solvation free energy of hydrophobic solutes in water by an entropic penalty due to an ordered "iceberg" structure of water molecules that forms around the solute. The term "iceberg" is not meant to be taken literally, but rather refers to a higher ordering of the first few solvation shells compared to bulk water\cite{iceberg_model, Grabowska_2021}. 

Indeed, the hydrophobic effect has been shown to be mainly entropy-driven\cite{Chandler_2005} and such ordered structures have been found around hydrophobic solutes\cite{Head_1995, Noskov_2005, Galamba_2013, Grabowska_2021}. However, in seminal papers Ben-Naim\cite{Ben_1975, Ben_1984, Ben_2013} and Yu et al.\cite{Yu_1988} have analytically proven that for pairwise interactions, the entropic and the enthalpic parts of the water-water interactions exactly cancel and, therefore, do not contribute to the net free energy change of solvation. This important theorem has led to the general understanding that \textit{any} solvent-response to the solute, such as the Frank and Evans "icebergs", has a net-zero effect on the free energy change and that, therefore, the solvent response cannot drive solvation. However, ordered structures of water molecules around hydrophobic solutes are still implied as the cause of hydrophobicity\cite{Reynolds_2001, Snyder_2014}, a notion that has then been corrected by others\cite{Graziano_2014_comment, Ben_2013}. Overall, this seeming contradiction caused considerable confusion, and still does.   

Against this background, our recent finding that water-water correlations contribute markedly to the folding free energy of crambin\cite{Heinz_2021_crambin} bears the potential for further confusion, as indeed testified by anonymous referee reports. Their comments have prompted us to look deeper into the subject, and to share our analysis as well as two illustrative examples with a broader readership.

Specifically, in our previous paper\cite{Heinz_2021_crambin} we calculated and compared solvation shell entropies of the solvated folded conformation of crambin and a molten-globule-like conformation of this prototypic globular protein. To this end the method Per|Mut\cite{Heinz_2019_rotational, Heinz_2020_translational} was used, which employs a mutual information expansion\cite{Matsuda_2000,Hnizdo_2007,Hnizdo_2008,Fengler_thesis}
\begin{equation}
    T\Delta S \approx T\Delta S_1 + T\Delta S_{\geq 2}
    \label{eq:MIE}
\end{equation}
into single-molecule entropies $\Delta S_1$ and entropy contributions $\Delta S_{\geq 2}$ arising from correlations between pairs and triples of --- mostly nearby --- solvent molecules. In our molecular dynamics simulations, the molten-globule-like conformation of crambin showed many hydrophobic residues, which are buried within the folded conformation, exposed to the solvent. Relative to the native fold, we observed indeed a marked entropic free energy contribution to the solvation free energy due to strongly correlated water molecules in the innermost solvation shells. 

This result may seem to be incompatible with the Ben-Naim theorem, in which case the notion that the solvent response cannot be a thermodynamic driving force would need to be reconsidered.

Here we will show that this finding is in fact --- counter-intuitively --- perfectly compatible with the Ben-Naim theorem. Our analysis will, further, provide a deeper understanding of the contribution of the solvent response to the solvation free energy. We will illustrate our reasoning by a simple Ising model example that can be exhaustively enumerated as well as by a more realistic example of a Lennard-Jones particle solvated in liquid argon. 

\section{Theory}
\label{sec:theory}

\subsection{Canonical decomposition}
Ben-Naim \cite{Ben_1975, Ben_1984, Ben_2013} has proven that the change of average solvent-solvent interaction energies upon solvation is exactly compensated by a corresponding entropy change, such that there is no net free energy contribution. Later, Yu et al.\cite{Yu_1988} obtained essentially the same result by considering a solvation process described by the coupling parameter $\lambda$ ($\lambda = 0$: not solvated, $\lambda = 1$: fully solvated). 

In particular, they demonstrated that for a Hamiltonian 
\begin{equation}
    \mathcal{H}(\lambda) = \mathcal{H}_{uv}(\lambda) + \mathcal{H}_{vv}
    \label{eq:hamiltonian}
\end{equation}
consisting of pairwise solute-solvent ($uv$) and solvent-solvent ($vv$) interactions, the internal energy ($\Delta U$) and entropy ($\Delta S$) changes can be expressed as
\begin{subequations}
    \label{eq:deltaU_decomp}
    \begin{align}
        \Delta U &= \underbrace{\langle \mathcal{H}_{uv} \rangle_{\lambda=1}}_{\Delta U_{uv}} \nonumber \\
        &+ \underbrace{\beta\!\! \int_0^1\!\!\! d\lambda \left[ \langle \mathcal{H}_{vv} \rangle\! \left\langle \frac{\partial \mathcal{H}_{uv}}{\partial \lambda} \right\rangle_{\!\!\lambda}\!\! -\! \left\langle \mathcal{H}_{vv} \frac{\partial \mathcal{H}_{uv}}{\partial \lambda} \right\rangle_{\!\!\lambda} \right]}_{\Delta U_{vv}} \\
        &= \Delta U_{uv} + \Delta U_{vv},
    \end{align}
\end{subequations}
\begin{subequations}
    \label{eq:deltaS_decomp}
    \begin{align}
        T \Delta S &= \underbrace{\beta\!\!\int_0^1\!\!\!d\lambda\! \left[ \langle \mathcal{H}_{uv} \rangle\!\left\langle \frac{\partial \mathcal{H}_{uv}}{\partial \lambda} \right\rangle_{\!\!\lambda}\!\! -\! \left\langle \mathcal{H}_{uv} \frac{\partial \mathcal{H}_{uv}}{\partial \lambda} \right\rangle_{\!\!\lambda} \right]}_{T\Delta S_{uv}} 
        \nonumber \\
        & + \underbrace{\beta\!\!\int_0^1\!\!\!d\lambda\! \left[ \langle \mathcal{H}_{vv} \rangle\!\left\langle \frac{\partial \mathcal{H}_{uv}}{\partial \lambda} \right\rangle_{\!\!\lambda}\!\! -\! \left\langle \mathcal{H}_{vv} \frac{\partial \mathcal{H}_{uv}}{\partial \lambda} \right\rangle_{\!\!\lambda} \right]}_{T \Delta S_{vv}} \label{eq:Svv}\\
        &= T\Delta S_{uv} + T\Delta S_{vv}\,.
    \end{align}
\end{subequations}
Whereas the internal energy and entropy parts $\Delta U_{uv}$ and $T \Delta S_{uv}$ only contain solute-solvent interactions $\mathcal{H}_{uv}$, the remaining terms are referred to as "solvent-solvent" terms, $\Delta U_{vv}$ and $T \Delta S_{vv}$, respectively. The important finding by Ben-Naim and Yu et al.\ is that these two terms are identical and thus cancel in the net free energy difference
\begin{equation}
    \Delta F = \Delta U - T \Delta S = \Delta U_{uv} - T\Delta S_{uv},
    \label{eq:deltaF_decomp}
\end{equation}
which thus only contains the so called solute-solvent terms. 

Note, however, that the ensemble --- and thus \textit{all} averages, including the solvent-solvent terms --- are affected by both solvent-solvent {\em as well as} solute-solvent contributions. Due to this very fact, the terminology "solute-solvent" vs.\ "solvent-solvent" is highly misleading and, as will become clear further below, is the root of longstanding and widespread confusion.

In particular, the seeming absence of solvent-solvent terms in the free-energy balance has led to the widely held belief that the solvent response to the presence of a solute, e.g., solvent rearrangements such as the Frank and Evans "icebergs", cannot contribute as a thermodynamic driving force\cite{Ben_2013, Persson_2017_3d2pt}. 

\subsection{Mutual information expansion}
Additionally, because solvent-solvent terms affect the ensemble itself and therefore also --- implicitly --- contribute to $T \Delta S_{uv}$, it is also misleading to split entropic contributions, which are inherently non-pairwise, into contributions that are defined via pairwise interaction energies, as in equation~\ref{eq:hamiltonian}. As an alternative, and to gain physical insight into the solvation process that can be interpreted in a more straightforward manner, we suggest to use a mutual information expansion (MIE)\cite{Matsuda_2000,Hnizdo_2007,Hnizdo_2008,Fengler_thesis}, as, e.g., used in the recently developed method Per|Mut\cite{Heinz_2019_rotational, Heinz_2020_translational}. 

Accordingly, the total solvent entropy is decomposed into single-body entropies, akin to an ideal-gas term, and multi-body correlations, 
\begin{subequations}
\begin{align}
        S &\approx\!\sum_{i=1}^N S_1(i) 
            \!- \!\!\sum_{\substack{(j,k)\\\text{pairs}}} I_2(j,k) 
            \!+ \!\!\!\sum_{\substack{(l,m,n)\\\text{triples}}}  \!\! I_3(l,m,n) 
            \!+ \ldots \\
            &= S_1 + S_{\geq 2}\,,
    \label{eq:rot_MIE_trunc}
\end{align}
\end{subequations}
where $S_1(i)$ are the single-body entropies of the (three-dimensional) probability distributions of molecules $1, \hdots, N$; $I_2(j,k)$ and $I_3(l,m,n)$ are the two-body and three-body mutual information terms of molecule pairs and triples, respectively. These terms are defined as
\begin{subequations}
\label{eq:def_MI1_to_3}
\begin{align}
    I_2(j,k)    &=S(j)+S(k)-S(j,k) \label{eq:def_MI2} \\
    I_3(l,m,n)  &=
    \!\begin{aligned}[t]
        &S(l)+S(m)+S(n) \\
        &-S(l,m)-S(l,n)-S(m,n)\\
        &+S(l,m,n)
    \end{aligned} 
    \label{eq:def_MI3}
\end{align}
\end{subequations}
and represent the entropy change due to two and three-body correlations, respectively. In this notation, $S(j,k)$ and $S(l,m,n)$ are the entropies of the (six and nine-dimensional) marginal distributions of the full configuration space density $\varrho$ with respect to molecule pairs $j,k$ and triples $l,m,n$, respectively, e.g., 
\begin{equation}
    \varrho(j,k)= \int \varrho \prod_{\substack{p=1\\{p\neq j, p\neq k}}}^N d\bm{x}_p.
\end{equation}

A full MIE up to the $N$-body correlation term yields an exact entropy decomposition. In our numerical approach below, evaluation of the respective integrals would require sampling over the full $3N$-dimensional configuration space, however, which is impractical. We therefore truncated the expansion after the three-body correlations to obtain a good approximation of entropy, neglecting higher-order terms. For short-ranged interactions, these have indeed been demonstrated to be small\cite{Rubi_2017}.

\section{Methods}
\label{sec:methods}

\subsection{Ising model}
\label{subsec:methods_ising}

To assess the solvent response to a solute (e.g., a protein), as sketched in Fig.~\ref{fig:systems}A, we first considered the simple $4 \times 4$ sub-critical Ising model sketched in Fig.~\ref{fig:systems}B. 
In this model, each spin $\sigma_{i,j}=-1,+1$ interacts with its nearest neighbors with an interaction strength $J=0.2$ under periodic boundary conditions. Here, the spins mimic a solvent with the four most center spins (shaded in red) interacting with an external field $\lambda$, which mimics the interaction with a solute. Note that, because the solute is described purely by these interactions, the Ising model depicted in Fig.~\ref{fig:systems}B does not contain any explicit solute degrees of freedom.

Accordingly, the Hamiltonian reads
\begin{subequations}
    \begin{align}
        \mathcal{H}(\bm{x}) &= \mathcal{H}_\textrm{vv}(\bm{x}) + \mathcal{H}_\textrm{uv}(\bm{x}) \\
         &= -J \sum_\textrm{n.n.} \sigma_{i,j} \sigma_{i',j'} - 4\lambda \sum_\textrm{shell} \sigma_{i,j},
    \end{align}
\end{subequations}
where the first sum runs over all nearest neighbors, and the second sum runs over the spins shaded in red (i.e., the "solvation shell", the solute is not shown in Fig.~\ref{fig:systems}B). The probability of each state $\bm{x} \in \bm{X} = [-1,+1]^{4 \times 4}$ reads
\begin{equation}
    P(\bm{x}) = \frac{1}{Z} e^{\frac{-\mathcal{H}(\bm{x})}{k_B T}},
\end{equation}
with the partition function $Z$ chosen such that $\sum_{\bm{x} \in \bm{X}} P(\bm{x}) = 1$.

The entropy $S$ and the average solvent-solvent interaction energy therefore read
\begin{align}
    S &= -k_B T \sum_{\bm{x} \in \bm{X}} P(\bm{x}) \log P(\bm{x})\,, \\
    U_{vv} &= \sum_{\bm{x} \in \bm{X}} P(\bm{x}) \mathcal{H}_{vv}(\bm{x}).
\end{align}

Following Yu et al. \cite{Yu_1988}, all solute-solvent and solvent-solvent entropy changes were calculated according to equation~\ref{eq:deltaS_decomp}. These values are therefore subject to a small integration error due to the required numerical integration, for which we used 251 discrete $\lambda$-intermediates. For the Ising model, all calculations were carried out with unitless energies, i.e., $k_B = T = \beta = 1$.

\subsection{Argon}
\label{subsec:argon_methods}
\subsubsection*{MD simulations}

To quantify the response of an argon-type liquid to a Lennard-Jones solute, two systems were simulated, an unsolvated system containing 512 argon-type atoms and a solvated system with an additional immobilized Van-der-Waals sphere as a 'solute'. All molecular dynamics (MD) simulations were carried out using the software package Gromacs~2020.6\cite{Gromacsa,Gromacsb,Gromacs4,Gromacs4.5,Gromacs2014} with a leapfrog integrator with a $2\,$fs time step. The Van-der-Waals parameters of argon were taken from the CHARMM36m force field\cite{Karplus,charmm36,charmm36m}. The Lennard-Jones\cite{LJ_potential} parameters (particle size $\sigma$ and potential depth $\epsilon$) for the model solute were chosen as twice as those of argon to enhance the statistical significance of average energy differences. All Van-der-Waals interactions were switched between $1.0\,$nm and $1.2\,$nm and no dispersion correction was applied. To immobilize the solute at the center of the simulation box, the freeze-options within Gromacs were used. During all simulation runs, the temperature was kept at $120\,$K using the V-rescale thermostat\cite{V_rescale} with a time constant of $0.1\,$ps. 

The unsolvated (pure argon) system was equilibrated at $1\,$bar pressure in a $20\,$ns NPT-run using the Berendsen barostat\cite{Berendsen_1984}, resulting in a $(3.073\,\textrm{nm})^3$ cubic simulation box. The second, solvated system was prepared by adding the solute to the simulation box and allowing for a further equilibration, lasting $10\,$ns under NVT conditions. For both systems, production runs, each lasting $4\,\mathrm{\mu}$s, were carried out under NVT conditions. For subsequent analysis, configurations were stored every $10\,$ps, resulting in trajectories consisting of $4\cdot 10^5$ frames each.

\subsubsection*{Entropy calculation}

Entropy contributions were calculated using the method Per|Mut\cite{Heinz_2019_rotational, Heinz_2020_translational}, which utilizes a permutation reduction\cite{Reinhard_2007_permute, Reinhard_2009_g_permute} and a mutual information expansion (MIE)\cite{Matsuda_2000,Hnizdo_2007,Hnizdo_2008,Fengler_thesis} into one-, two- and three-body correlations. For permutation reduction, 50 different simulation snapshots were randomly selected as reference structures and a MIE was carried out using each of the permutationally reduced trajectories. In the MIE, the mutual information between all pairs of argon atoms was taken into account; triple-wise mutual information terms were cut off at an average distance of $0.5\,$nm after permutation reduction\cite{Heinz_2019_rotational, Heinz_2020_translational}. All MIE orders were calculated using a k-nearest-neighbor algorithm with a value of $k=1$.

From the resulting entropy difference $\Delta S_\textrm{MIE}$ between the unsolvated system and the solvated system, the free energy difference $\Delta F_{\textrm{MIE}} = \Delta U - T \Delta S_{\textrm{MIE}}$ was calculated, where the internal energy difference $\Delta U$ was obtained directly from the average interaction energies in the simulation runs. 

Solute-solvent and solvent-solvent entropy differences $\Delta S_\textrm{uv}$ and $\Delta S_\textrm{vv}$, respectively, were calculated by thermodynamic integration (TI) from the unsolvated state to the solvated state using 200 equidistant windows, each lasting $200\,$ns. As a control for the Per|Mut results, the total entropy difference $\Delta S_{\textrm{TI}} = \Delta S_\textrm{uv} + \Delta S_\textrm{vv}$ and the free energy change 
\begin{equation}
    \Delta F_{\textrm{TI}} = \int_0^1 \left\langle  \frac{\partial \mathcal{H}}{\partial \lambda} \right\rangle d\lambda
\end{equation}
were calculated using standard TI.

Errors of the internal energies were calculated as $\sigma_U/\sqrt{N_f-1}$, where $\sigma_U$ are the standard deviations of the respective interaction energies from $N_f=400 \cdot 10^3$ simulation frames. Due to the long interval of $10\,$ps between frames, these were considered statistically independent.
Similarly, Per|Mut errors were estimated as the standard errors resulting from the 50 permutationally reduced simulation trajectories. 
TI errors were estimated from the difference between two independent sets of TI simulation runs with identical input parameters but different initial (random) velocities, but turned out to be negligible for all further analyses. 

\section{Results and discussion}


To investigate the seeming contradiction between the Ben-Naim theorem and the free energy effects of increased solvent correlations observed for the crambin\cite{Heinz_2021_crambin} solvent shell, we calculated the relevant contributions of the solvent response to the solvation free energy for two simple model systems, for which sampling errors can be neglected. Specifically, we will compare all free energy contributions of the Yu et al.\ decomposition (equations~\ref{eq:deltaU_decomp}-\ref{eq:deltaF_decomp}) with the mutual information expansion (equation~\ref{eq:MIE}). 

We will first consider an idealized solvation process for an Ising model, for which all relevant quantities can be exhaustively enumerated, such that the results are exact to numerical precision. Subsequently, we will consider a liquid argon-type Lennard-Jones system with a van-der-Waals solute, the relaxation times of which are short with respect to simulation times, such that for this more realistic system sampling errors can be assumed to be very small with respect to the relevant energy and entropy differences.

\subsection{Ising model}
\label{subsec:ising_results}

\begin{figure}[h!]
    \centering
    \includegraphics[width=7.5cm]{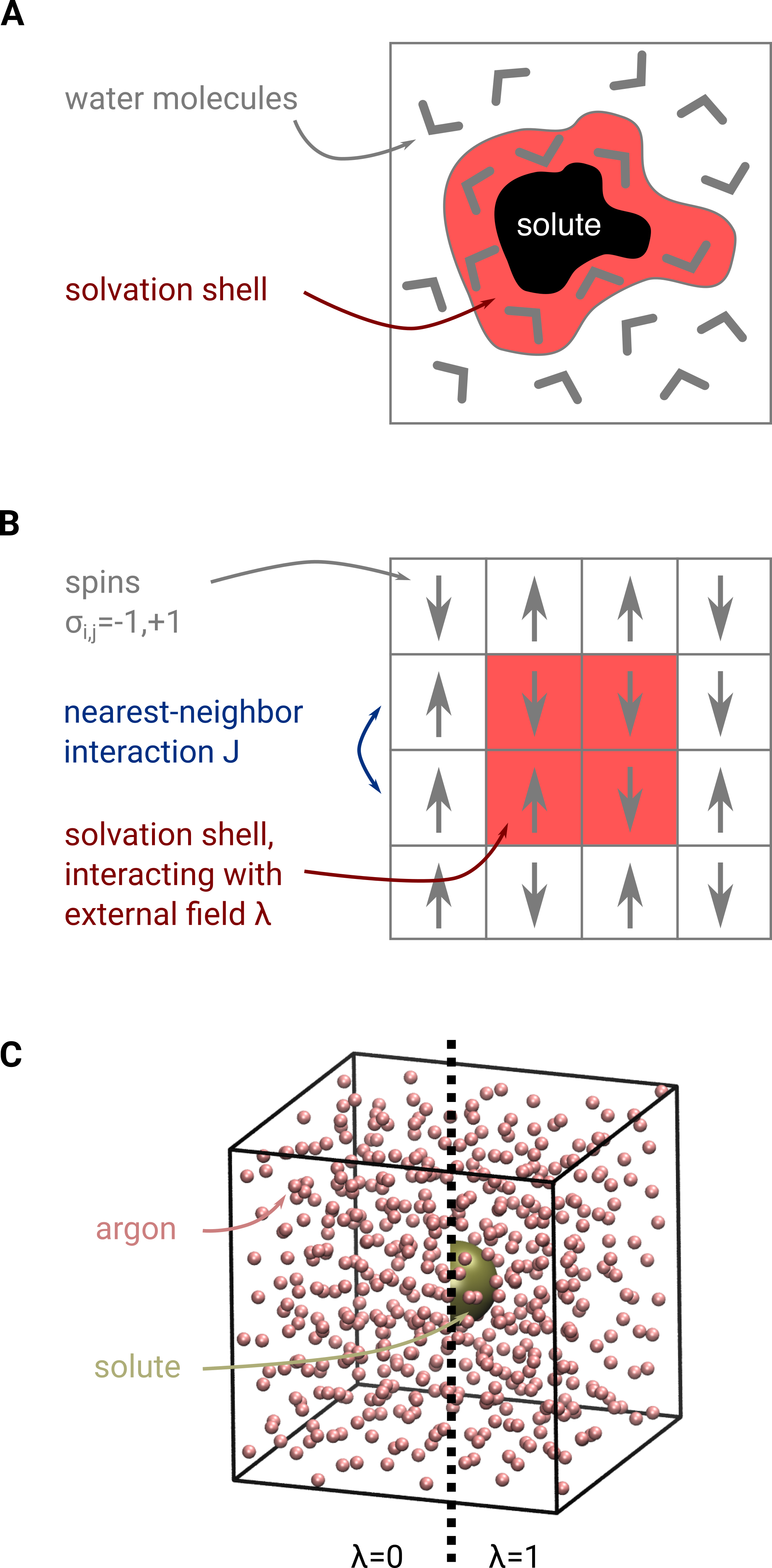}
    \caption{(A) Sketch of a solute (black) in water (gray angles), where the solvation shell (red area) interacts directly with the protein. (B) Sketch of a $4 \times 4$ Ising model, serving as a simplified model of solute-solvent / solvent-solvent interactions. Here, spins (gray arrows) represent water molecules. The four spins shaded in red interact with the solute; this interaction is described by an external field $\lambda$ that acts on the four spins. (C) As a more realistic model system, argon-type atoms (red) are coupled to a solute Lennard-Jones sphere (yellow). On the left ($\lambda = 0$), the solute is decoupled from the argon solvent; on the right ($\lambda = 1$), the solute is fully solvated.}
    \label{fig:systems}
\end{figure}

As a simple illustrative model of a solute in a solvent (Fig.~\ref{fig:systems}A), we consider the $4 \times 4$ sub-critical Ising model shown in Fig.~\ref{fig:systems}B. Here, each spin represents a solvent molecule that interacts with its nearest neighbors. The effects of a solute are modeled by an external field with strength $\lambda$ that acts on the "solvation shell" (red), consisting of the four spins at the center.

\begin{figure}[h!]
    \centering
    \includegraphics[width=8cm]{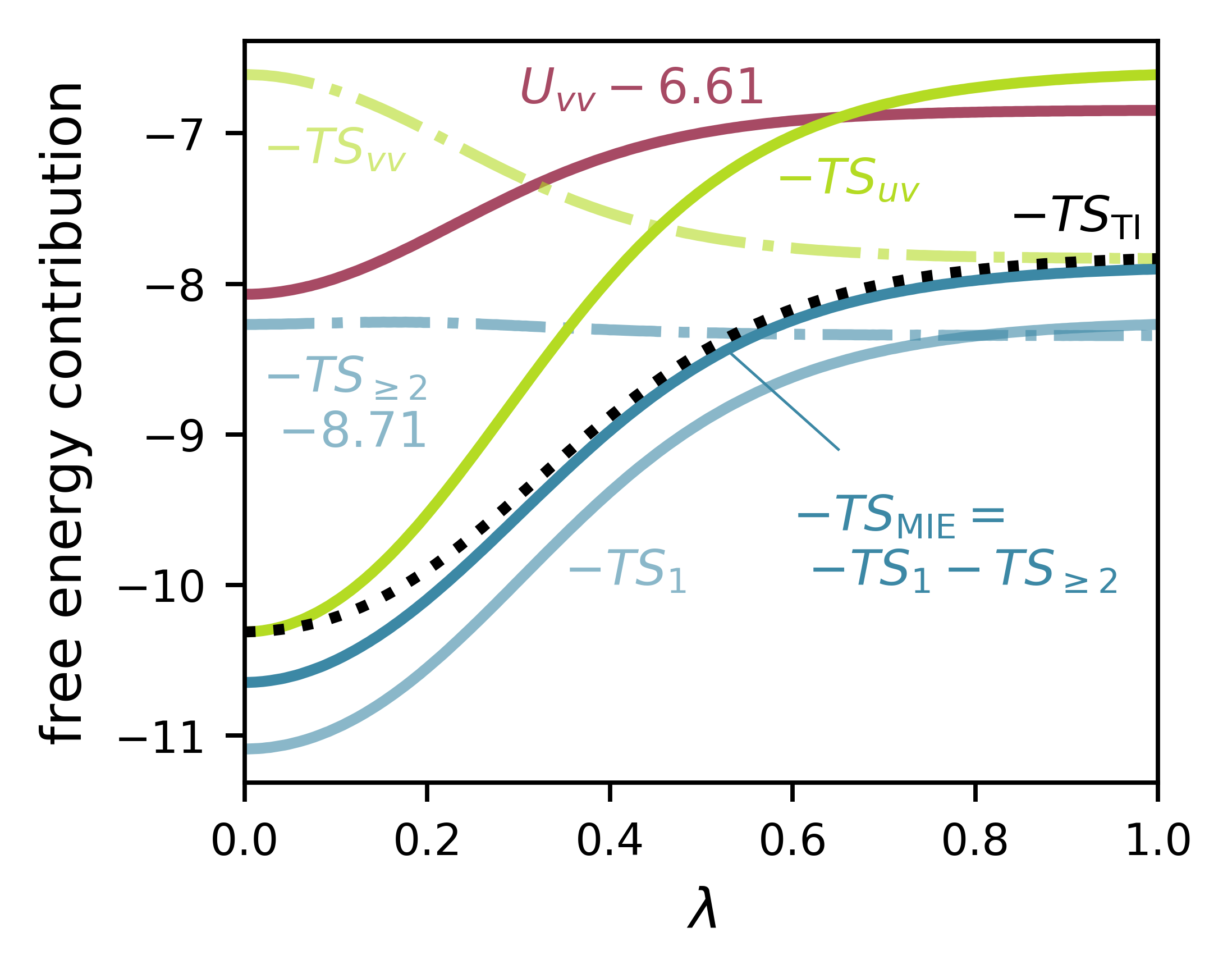}
    \caption{Thermodynamic quantities of the Ising model as a function of the external field ($\lambda$). The precise entropy contribution ($-TS_\textrm{TI}$) is shown as a dotted black line. The single-body entropy ($-TS_1$) from the Per|Mut MIE is shown as a solid light blue line; the contribution from two- and three-body correlations ($-T S_{\geq 2}$) is shown as a dashed-dotted light blue line. The entropy from the MIE approximation, including correlations, is shown in dark blue. The entropy decomposition into $-TS_{uv}$ (solid line), and $-TS_{vv}$ (dashed-dotted line) is shown in green. The red line represents the solvent-solvent interaction energies ($U_{vv}$). For a better visual representation, $U_{vv}$ and $-TS_{\geq 2}$ are shifted by $6.61$ and $8.71$ units, respectively.}
    \label{fig:ising}
\end{figure}

Figure~\ref{fig:ising} shows the exact relevant thermodynamic quantities as a function of the coupling parameter $\lambda$, calculated by full enumeration as described in section~\ref{subsec:methods_ising}. As expected, with increasing coupling to the solvent, the total entropic free energy contribution $-T S_\textrm{TI}$ (dotted black line) becomes less favorable (i.e., it increases), and eventually saturates at an entropy difference of $-T \Delta S_\textrm{TI} = 2.48$ between fully solvated ($\lambda=1$) and fully decoupled ($\lambda=0$). 
This contribution is dominated by the unfavorable solute-solvent contribution $-T \Delta S_{uv} = 3.70$ (green solid line), which is partially compensated by the favorable solvent-solvent contribution $-T \Delta S_{vv} = -1.22$ (green dashed-dotted line). 

As shown by the solid red line in Fig.~\ref{fig:ising}, the average solvent-solvent interaction energies $U_{vv}$ increase for increasing coupling parameter $\lambda$ and thus contribute unfavorably to the free energy change by $\Delta U_{vv} = 1.22$. Fully in line with the Ben-Naim theorem (equation~\ref{eq:deltaF_decomp}), $U_{vv}$ is indeed precisely compensated by $-TS_{vv}$, such that $\Delta U_{vv}-T\Delta S_{vv}=0$  and, hence, $\Delta F=\Delta U_{uv} - T \Delta S_{uv}$ ($\Delta U_{uv} = -15.98$, not shown in Fig.~\ref{fig:ising}).

Does this finding imply that solvent-solvent correlations do not contribute to the solvation free energy? To answer this question, consider the above mutual information expansion of entropy, which directly quantifies these correlations. 
The single-body term $-TS_1$ (light blue line) underestimates the entropy on average by $0.53$ units and contributes $-T\Delta S_1 = 2.82$ to the overall entropy change. Inclusion of the two- and three body correlation terms ($-T \Delta S_\textrm{MIE}$, solid blue line) improves the approximation  markedly, with an average deviation from the exact values below $0.13$ units. The correlation terms $-T S_{\geq 2}$ (dashed-dotted light blue line) add a small favorable contribution of $-T \Delta S_{\geq 2} = -0.07$ to the overall MIE entropic free energy change of $-T \Delta S_\textrm{MIE} = 2.75$.

Crucially, the term $-T S_{vv}$ differs from the solvent-solvent correlations $-TS_{\geq 2}$ both by definition and, indeed, also numerically as shown in Fig~\ref{fig:ising}. As a result, also $-T S_{uv}$ differs from $-T S_1$, and the entropy change due to solvent correlations $-T \Delta S_{\geq 2}$ is also not compensated by any canonical internal energy term. 
This simple example illustrates that, generally, solvent correlations do contribute to the solvation free energy; it also clarifies why this finding is not in conflict with the Ben-Naim theorem.


\subsection{Argon}
Is this subtle but important distinction between $-TS_{vv}$ and the actual many-body contribution to the solvation entropy, $-TS_{\geq 2}$, also relevant for more realistic systems? To address this question, we carried out MD simulations of a system comprising 512 argon-type atoms and an immobilized Lennard-Jones "solute", as described in section~\ref{subsec:argon_methods} (see also  Fig.~\ref{fig:systems}C). Here we calculated the free energy change of solvation, as well as the relevant enthalpic and entropic contributions using both, Per|Mut and thermodynamic integration.

\begin{figure}[h!]
    \centering
    \includegraphics{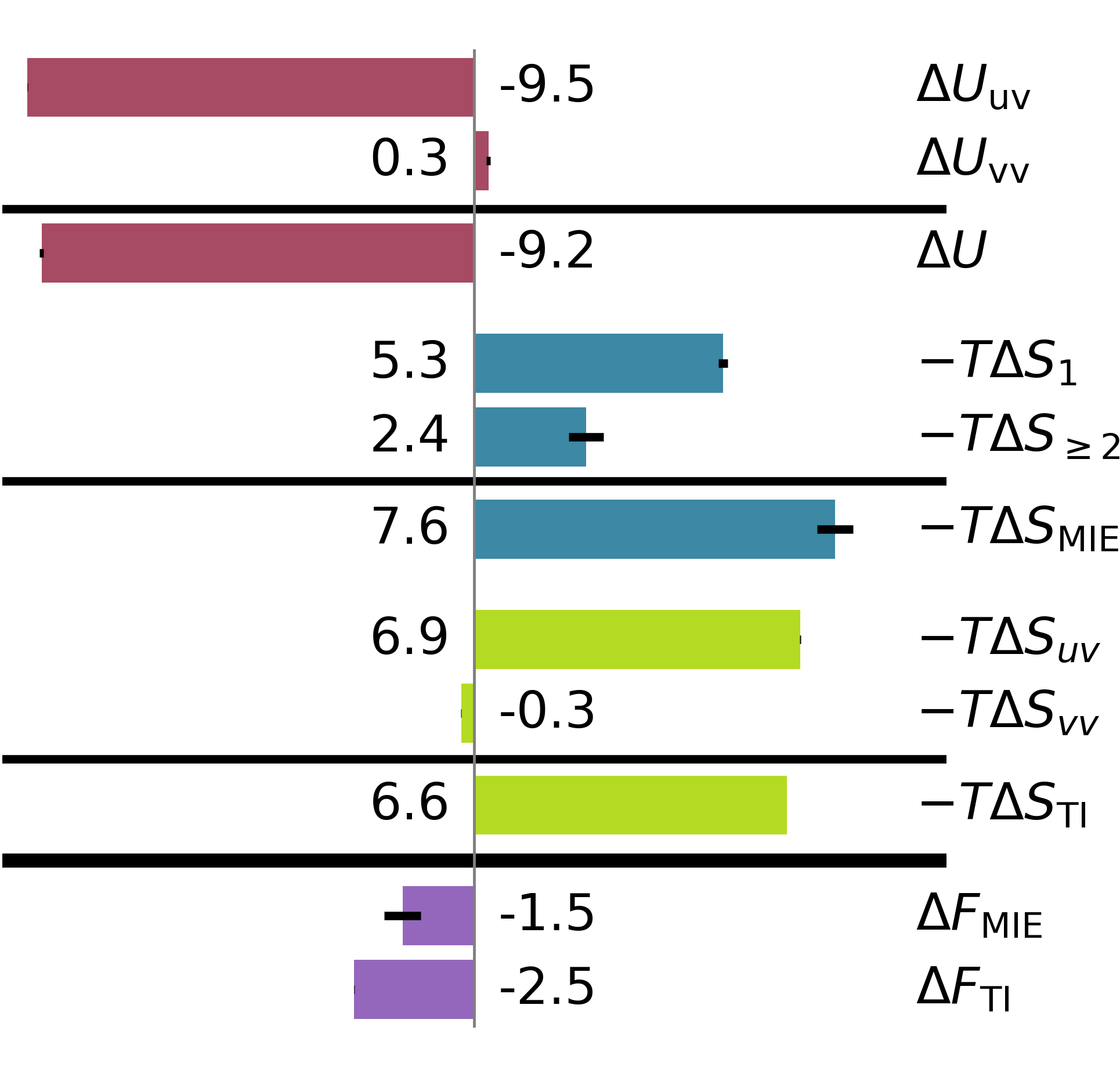}
    \caption{Solvation free energy contributions (in \kJmol{}) of the fixed Lennard-Jones solute in an argon-type liquid. Red bars denote the internal energy change and its contributions. Blue and green bars denote the entropy change and its contributions, calculated using Per|Mut and TI, respectively. Purple bars show the overall free energy change, as calculated using Per|Mut and TI. Estimated sampling uncertainties are shown as small black bars.}
    \label{fig:argon}
\end{figure}

As shown in Fig.~\ref{fig:argon}, the internal energy change $\Delta U$ upon solvation is favorable and totals $-9.2\,$\kJmol{}, to which solvent-solute interactions ($\Delta U_{uv}$) contribute $-9.5\,$\kJmol{} and solvent-solvent interactions ($\Delta U_{vv}$) contribute $0.3\,$\kJmol{}. In line with the Ben-Naim theorem, the later contribution is exactly compensated by $-T \Delta S_{vv} = -0.3\,$\kJmol{}, which, also for this system, might suggest that the solvent-solvent interactions and correlations, taken together, do not contribute to the solvation free energy. 

However, the many-body entropy contribution $-T \Delta S_{\geq 2} = (2.4 \pm 0.4)\,$\kJmol{}, calculated using Per|Mut and the mutual information expansion, is substantial and contributes a significant fraction to the solvation entropy $-T \Delta S_\textrm{MIE} = (7.6 \pm 0.4)\,$\kJmol{}, which is dominated by the reduced volume of the individual argon atoms, $-T \Delta S_{1} = (5.3 \pm 0.1)\,$\kJmol{}.

To test our assumption that four-body and higher correlations not included within $-T \Delta S_\textrm{MIE}$ are sufficiently small, we have also calculated the relevant entropy terms using TI (green). Indeed, the similar total entropy change of $-T \Delta S_\textrm{TI} = 6.6\,$\kJmol{} supports this assumption and shows that the contribution of the higher correlations to the solvation entropy is markedly smaller than the MIE estimate. Also for this more realistic system, the entropy change due to solute-solvent interactions ($-T \Delta S_{uv} = 6.9\,$\kJmol{}) dominates, and $-T\Delta S_{vv}$ does not even describe the correct sign of the actual solvent-solvent correlation contribution to the solvation free energy.
The remaining difference of ca.\ $1\,$\kJmol{} between the MIE and TI solvation entropies is also reflected in the respective total free energies $\Delta F_\textrm{MIE}=\Delta U - T \Delta S_\textrm{MIE} = -1.5\,$\kJmol{} and $\Delta F_\textrm{TI} = -2.5\,$\kJmol{}, respectively, underscoring that this difference is mainly due to the truncated MIE expansion rather than sampling uncertainties.

Similar to our findings for the above Ising model, also for the more realisic argon-type system the two possible entropy decompositions differ significantly. Whereas the small size of the two solvent-solvent terms $\Delta U_{vv}$ and 
$-T\Delta S_{vv}$ --- and in particular their mutual cancellation --- seem to show that the solvation of this Lennard-Jones particle is unaffected by the reaction of the solvent, the actual solvent-solvent entropy contributions are substantial and not compensated by any canonical internal energy term. 
We conclude that also for the solvation of a Lennard-Jones particle in a Lennard-Jones fluid, the induced solvent reorganization contributes markedly to the solvation free energy.

\section{Conclusions}
We pointed out that the entropy decomposition by Ben-Naim and Yu et al.\ into a contribution $S_{uv}$ from solute-solvent interactions and a remaining contribution ($S_{vv}$) as defined in equation~\ref{eq:deltaS_decomp}, differs conceptually from direct evaluation --- e.g., via a mutual information expansion --- of solvent-solvent correlation contributions to the solvation free energy. 
In particular, the term "solvent-reorganization entropy" for $\Delta S_{vv}$ is highly misleading, because it creates the wrong impression that any solvent response to the presence of a solute cannot contribute to the net free energy.

Two examples served to illustrate the solution of this seeming paradox. First, a simple semi-analytical Ising model, which permitted exhaustive enumeration, establishes that the conceptual difference between $\Delta S_{vv}$ and $\Delta S_{\geq 2}$ actually gives rise to marked numerical differences. Second, our MD simulations of solvation within a Lennard-Jones liquid 
show that this distinction is also relevant for a more realistic solvation system. For both systems, $\Delta S_{vv}$ is exactly compensated by the change of average solvent-solvent interactions ($\Delta U_{vv}$), as required by Ben-Naim's theorem.

In more general terms --- as already pointed out by Lee\cite{Lee_1994} --- one can define for any change of an entropy component $T \Delta S_{xy}$ an appropriate $\Delta U_{xy}$ component such that $T \Delta S_{xy} = \Delta U_{xy}$. However, such construction does not necessarily allow for a physically meaningful interpretation; in particular, it does not support the conclusion that '$xy$' is irrelevant for the solvation process. Whereas the canonical decomposition of pairwise interaction energies into solvent-solute ($\mathcal{H}_{uv}$) and solvent-solvent ($\mathcal{H}_{vv}$) terms, as well as the corresponding decomposition of internal energies, are certainly physically meaningful, this does not necessarily apply to the corresponding entropy terms due to their inherently non-pairwise nature. Instead, an entropy decomposition into a single-body term and multi-body correlations provides a more intuitive understanding.

We hope our explanations and examples will contribute to resolving a long-standing controversy and the resulting widespread confusion. Fully in line with Ben-Naim's theorem, solvent-solvent correlations can --- and generally do --- contribute markedly to the overall free energy of solvation, thus underscoring the need for an improved understanding of the "iceberg"-type ordering of solvent shells, in particular near complex macromolecular solutes and surfaces. 

\begin{acknowledgement}

We thank the anonymous referees of our previous paper\cite{Heinz_2021_crambin} for pointing out this seeming contradiction, which triggered the present analysis, and Petra Kellers for proofreading the manuscript. 

\end{acknowledgement}

%
%

\bibliography{references}

\clearpage

\end{document}